\documentclass[preprint,12pt]{elsarticle}

\usepackage{amssymb}
\usepackage{algpseudocode} 
\usepackage{algorithm, float} 
\usepackage{multirow}
\usepackage{svg} 
\DeclareUnicodeCharacter{2212}{\ensuremath{-}} 

\journal{arXiv}

\patchcmd{\pprintMaketitle}
 {\hrule}
 {\clearpage\hrule}
 {}{}
\appto\endfrontmatter{\clearpage}

\begin{document}

\begin{frontmatter}

\title{Agile gesture recognition for low-power applications: customisation for generalisation \thanks{The project was supported by the Innovate UK KTP grant (12250) and by the UKRI Turing AI Acceleration Fellowship EP/V025295/1, EP/V025295/2.}}


\author[inst1,inst2]{Ying Liu}
\affiliation[inst1]{organization={University of Leicester},
            addressline={University Rd}, 
            city={Leicester},
            postcode={LE1 7RH}, 
            country={U.K.}}
\author[inst2]{Liucheng Guo}
\affiliation[inst2]{organization={Tangi0 Ltd.},
            addressline={10 Lion Yard}, 
            city={London},
            postcode={SW4 7NQ}, 
            country={U.K.}}
\author[inst3]{Valeri A. Makarov}
\affiliation[inst3]{organization={Interdisciplinary Mathematics Institute, Universidad Complutense de Madrid},
            addressline={Plaza de las Ciencias 3}, 
            city={Madrid},
            postcode={28040}, 
            country={Spain}}
\author[inst1]{Alexander Gorban}
\author[inst1]{Evgeny Mirkes}
\author[inst1,inst4]{Ivan Y. Tyukin}
\affiliation[inst4]{organization={King's College London},
            addressline={Strand}, 
            city={London},
            postcode={WC2R 2LS}, 
            country={U.K.}}
\ead{ivan.tyukin@kcl.ac.uk}

\begin{abstract}


    Automated hand gesture recognition has long been a focal point in the AI community. Traditionally, research in this field has predominantly focused on scenarios with access to a continuous flow of hand's images. This focus has been driven by the widespread use of cameras and the abundant availability of image data. However, there is an increasing demand for gesture recognition technologies that operate on low-power sensor devices. This is due to the rising concerns for data leakage and end-user privacy, as well as the limited battery capacity and the computing power in low-cost devices. Moreover, the challenge in data collection for individually designed hardware also hinders the generalisation of a gesture recognition model.
    
    In this study, we unveil a novel methodology for pattern recognition systems using adaptive and agile error correction, designed to enhance the performance of legacy gesture recognition models on devices with limited battery capacity and computing power. This system comprises a compact Support Vector Machine as the base model for live gesture recognition. Additionally, it features an adaptive agile error corrector that employs few-shot learning within the feature space induced by high-dimensional kernel mappings. The error corrector can be customised for each user, allowing for dynamic adjustments to the gesture prediction based on their movement patterns while maintaining the agile performance of its base model on a low-cost and low-power micro-controller. This proposed system is distinguished by its compact size, rapid processing speed, and low power consumption, making it ideal for a wide range of embedded systems.

\end{abstract}

\clearpage

\begin{keyword}
few-shot learning \sep customised gesture recognition \sep embedded system
\end{keyword}

\end{frontmatter}


\section{Introduction}
\label{introduction}

    Hand gesture recognition algorithms have seen intensive and rapid development in recent years, driven by technological advancements and the increased availability of personal camera devices \cite{Oudah2020review}. Past research mainly focused on recognising specific gestures from vision-based systems, which use advanced algorithms to detect hand gestures from image data \cite{Guo2021reviewHMI}. However, the high-dimensional image data in the gesture recognition system requires extensive computing capability unsuitable for low-power, low-cost devices. The live recognition event also requires a fast response time, making it challenging for vision-based systems due to the long interpretation time of image data.  Additionally, vision-based systems face the risk of accidental or adversarial leakage of personal information.

    Another approach to recognising hand gestures is hardware-based embedded systems, which measure signals from muscle movements and classify them. It shows potential for rapid signal analysis, as opposed to interpreting high-dimensional image data, especially in scenarios like human-computer interaction, human behaviour analysis, and accessibility solutions for people with movement disorders \cite{Benalcazar2017Myo, Wang2020GR, Moin2020wearable, Oweis2011EEG}. Hardware-based systems also significantly reduce the risk of accidental or adversarial leakage of personal information, as they do not require video or photographic imagery for gesture acquisition. However, there are difficulties in collecting large amounts of training data with a wide variety of users for specific hardware. On top of that, each user has unique finger movement patterns, body size, and different gesture habits in real life, which is challenging to have a generalised solution for the gesture recognition \cite{Lobov2018}. 
    
     Most hand gesture recognition algorithms use neural networks, as they effectively classify high-dimensional data, such as surface electromyography signals \cite{T.2023} or images\cite{Oudah2020review}, or . However, they are not applicable for live gesture recognition systems in low-power, low cost devices due to the high-power and slow processing time. The difficulties in data collection of hardware-based systems also made it challenging to train a gesture recognition neural network for generalisation purposes.

    In this work, we propose a new method to address the challenges of long processing time, high computing power, limited training data and data privacy risks of the existing gesture recognition systems. The method is based on a combination of an agile and fast classical support vector machine (SVM) model equipped with an "adaptation add-on" capable of fine-tuning the model for the end-user. It utilises the few-shot learning and "blessing of dimensionality" to distinguish errors from the SVM model and correct them to the right gesture according to the individual user's pattern. To demonstrate this capability, we used the hand controller \textit{etee} to record hand movement signals via capacitive sensors \cite{etee}. We collected over 20,000 tactile frames from 12 users performing 4 types of dynamic hand gestures. Using this tactile information, we designed and implemented a gesture classifier with an adaptive error correction mechanism. This system is distinguished by its compact size, high processing speed, and minimal power consumption. It is ideal for  a wide range of embedded systems, including the \textit{etee} with its 0.85 W power usage.

    The paper is organised as follows: Section \ref{sec:method} introduces the methodology of the error correction system. Section \ref{sec:Hardware} presents the hardware architecture, the hand gestures used in the dataset, and the data collection and reprocessing protocols. Section \ref{sec:results} details the error correction results, including the base multi-classification models for distinguishing dynamic gestures, error data analysis, and the performance of our adaptive error corrector. Section \ref{sec:conclusion} concludes the paper.


\section{The Method}\label{sec:method}

    At the heart of our work is the inherent capability of algorithms to adapt to change. This includes the capability to learn in weakly supervised fashion from few labelled data instances. 
    
    The recent study by Sutton et al. \cite{Sutton2022a} highlights the advantages of nonlinear feature mappings in few-shot learning, particularly when the high-dimensional features exhibit a significant degree of orthogonality. In contexts where the set of errors and the set of correct predictions by the base model are substantially unbalanced, this property can be effective. The high degree of orthogonality in the high-dimensional feature space facilitates the separation of base model prediction errors from correctly predicted instances, thereby enhancing the accuracy and reliability of the model in scenarios with limited data availability.

    The error corrector, developed by Tyukin and Gorban, leverages the principles of the concentration of measure and the stochastic separation theorem in high dimensions \cite{error_corrector_2018}. The foundational concept, derived from the classical concentration of measure theorem, suggests that independent and identically distributed (i.i.d.) random points in high-dimensional spaces tend to congregate on a thin layer of a sphere's surface \cite{concentration_measure_1994}. Further, these points are linearly separable from each other, as established in their works \cite{stochastic2017, stochastic2021}. This implies that in high-dimensional spaces, errors can be linearly separated from the rest of the samples, a phenomenon that has been effectively utilised in various applications. For instance, the performance monitoring of computer numerical control milling processes and edge-based object detection \cite{corrector_application2021}. This characteristic of high-dimensional spaces is often referred to as the "blessing of dimensionality."
    
    \subsection{General overview of the method's workflow}
    
    Regardless of the gesture recognition model, errors are inherent in the inference process. When we introduce data from a new user, variations in gesture patterns arise due to differences in hand size and personal habits. This variability means the model cannot generalise effectively for all users, leading to increased recognition errors with new user data. To enhance the model's adaptability for new users while maintaining its original performance, we incorporate an error corrector into the gesture recognition system, utilising the "blessing of dimensionality". 
    
    The process begins with the input sensor signals processed by a base model for gesture recognition. Subsequently, these features are also mapped into a high-dimensional feature space using a kernel map. Finally, a centroid classifier, serving as a separator, is applied within this high-dimensional space to distinguish the base model errors. Finding the error does not necessarily mean that we know the correct gesture. We also need to find the type of errors. Hence, an error-type classifier needs to be trained with the error data to direct to the correct gesture.  

\begin{figure*}
        \includegraphics[width=1\textwidth]{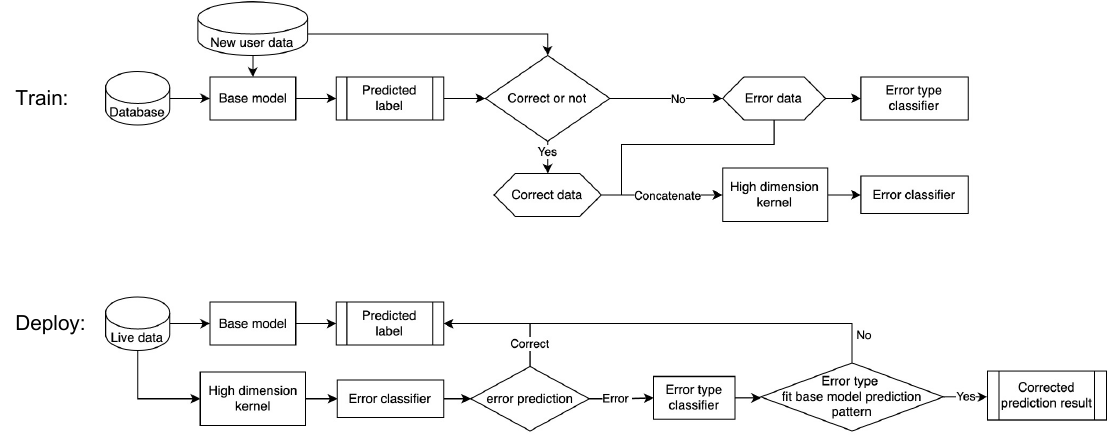}
        \caption{The flowchart of the adaptive error corrector.}
        \label{fig:workflow}
    \end{figure*}

    Figure~\ref{fig:workflow} outlines the complete training and deployment procedures for the gesture recognition system. The procedure commences with the base model processing the input sensor signals. These data are then projected into a high-dimensional feature space via a kernel map. A centroid classifier, functioning as a separator, is subsequently employed in this high-dimensional space to discern errors made by the base model. Since the base model operates as a multi-classification model, identifying an error does not automatically provides the correct gesture. Therefore, it becomes necessary to classify the types of errors. For this, an error-type classifier is trained using the error data to guide toward the accurate gesture.

\subsection{Training and Deployment Algorithms }

    During training the base model, a new dataset from a new user is introduced (Figure~\ref{fig:workflow} top, see also Algorithm~\ref{algo: train}). The predicted label is compared with the actual gesture to generate an error/correct label. The error instances in this dataset are then utilised to train an error-type classifier. Each type corresponds to a scenario where the true label is $A$ and the base model's prediction is $B$. Concurrently, the new user dataset is mapped to a high-dimensional feature space using a kernel map. The high-dimensional features, and their correct/error labels, are segregated using an error corrector like the centroid classifier. The high-dimensional kernel, the error-type classifier, and the error corrector are embedded in the hardware for error correction during deployment.

    In the deployment phase (Figure~\ref{fig:workflow} bottom, see also Algorithm~\ref{algo: deployment}), the base model processes live data streamed from sensors to predict a gesture. Simultaneously, it is projected into the high-dimensional feature space and inputted into the error classifier. If the label is identified as correct, the base model's prediction is outputted as the final result. If classified as an error, the error-type classifier examines the error type. Each error type comprises two parts ($A$, $B$), with $A$ being misrecognised as $B$ in the base model. If the base model's prediction is not $B$, the error does not conform to the identified error type, and the original output is retained. Conversely, if the prediction is $B$, it aligns with the error type pattern, and the final output is corrected to $A$.

     \begin{algorithm}[H]
    	\caption{Error corrector with nonlinear feature maps: Training} 
        \label{algo: train}
    	\begin{algorithmic}[1]
             \Require Set $\mathcal{X}$ containing the correct dataset from a new user, and set $\mathcal{X}^*$ consisting of the error dataset specific to the new user. 
             \State Project $\mathcal{X}$ and $\mathcal{X}^*$ to a high dimensional space using nonlinear kernel map $\phi$ to construct dataset in the feature space $\phi(x), x\in(\mathcal{X}, \mathcal{X}^*)$. 
             \State Determining the centroid of the $\phi(x), x\in\mathcal{X}$ as $\mathbf{\bar{x}}$ . Generate two sets, the centralised correct set $\mathcal{X}_c = \phi(x) - \bar{\mathbf{x}}, x\in\mathcal{X}$ and error set  $\mathcal{X}_c^* =  \phi(x) - \bar{\mathbf{x}}, x\in\mathcal{X^*}$.
             \State For the centralised feature $\mathcal{X}_c$, apply Principal Component Analysis (PCA), $H = (h_1, ..., h_n)$, where $h_i$ are eigenvectors corresponding to the eigenvalues $\lambda_1 \geq \cdots \geq \lambda_n > 0$ of the covariance matrix of the set $\mathcal{X}_c$, and $n$ is the number of features in a sample.
             
                $\mathcal{X}_r = \{\mathbf{z}|\mathbf{z} = H\mathbf{x}, \mathbf{x}\in\mathcal{X}_c\}$
             
                $\mathcal{X}_r^* = \{\mathbf{z}|\mathbf{z} = H\mathbf{x}, \mathbf{x}\in\mathcal{X}_c^*\}$ 
             \State Apply whitening transformation to set $\mathcal{X}_c$ and $\mathcal{X}_c^*$ $W = diag(\frac{1}{\sqrt{\lambda_1}}), \cdots, \frac{1}{\sqrt{\lambda_n}}$.
             
                $\mathcal{X}_w = \{\mathbf{u}|\mathbf{u} = W\mathbf{z}, \mathbf{z}\in\mathcal{X}_r\}$
             
                $\mathcal{X}_w^* = \{\mathbf{u}|\mathbf{u} = W\mathbf{z}, \mathbf{z}\in\mathcal{X}_r^*\}$ 
            \State Construct centroid classifier with the normalised error centre $\hat{\mathbf{u}}^*=\bar{\mathbf{u}}^*/||\bar{\mathbf{u}}^*||, \mathbf{u} \in \mathcal{X}_w^*$, a decision boundary limits $\theta_{min}$ and $\theta_{max}$, and a threshold $\Delta$ that decision boundary $\theta = \theta_{min} + \Delta(\theta_{max} - \theta_{min})$ .
            \State A sample $v$ belongs to the correct set $\mathcal{X}$ \textbf{if} 
            \Statex \quad $(\hat{\mathbf{u}}^*,WH(\phi(v)-\bar{\mathbf{x}})) < \theta$,
            \Statex and it is in the error set $\mathcal{X}^{*}$ \textbf{otherwise}
            \State Construct Linear Discriminant Analysis (LDA) classifier $f$ to classify error type $\sigma = f(x), x\in(\mathcal{X}, \mathcal{X}^*)$, where $\sigma$ refers to a type of error $\sigma = (a,b)$ with true label $y = a$ and base model prediction $y_{pred} = b$. 
    		
    	\end{algorithmic} 
    \end{algorithm}

    \begin{algorithm}[H]
    \caption{Error corrector with nonlinear feature maps: Deployment} 
        
        \label{algo: deployment}
        \begin{algorithmic}[1]
           \Require A data vector $\mathbf{v}$, the centroid of the correct set in the training dataset $\bar{\mathbf{x}}$,  nonlinear feature mapping $\phi$, Principal Component transformation $H$ and whitening transformation matrix $W$, selected feature index $m$, normalised error centre $\hat{\mathbf{u}}^*$, threshold $\Delta$, decision boundary $\theta$ of the centroid classifier that $\theta = \theta_{min} + \Delta(\theta_{max} - \theta_{min})$, and error group classifier $f$.
           
        \State Compute centralised high dimensional features $\phi(\mathbf{v}) - \mathbf{\bar{x}}$.
        \State Apply PCA and whitening transformation $\mathbf{v}_w = WH(\phi(\mathbf{v}) - \mathbf{\bar{x}})$.
        \State Associate the vector $\mathbf{x}$ with the error set and determine error label $\epsilon$,
        \Statex \quad \textbf{if} $(\hat{\mathbf{u}}^*, \mathbf{v}_w) \geq \theta$:
        \Statex \quad \quad $\epsilon = 1$, it is an error,
        \Statex \quad \textbf{else}:
        \Statex \quad \quad $\epsilon = 0$, it is not an error.
        \State Find error type $\mathbf{\sigma} = f(\mathbf{v}), $, where $\mathbf{\sigma}$ refers to a type of error $(a,b)$.
        \State Construct the label when necessary.
        \Statex \textbf{if} $\epsilon = 1$ and $\mathbf{y}_{pred} = b$:
        \Statex \quad $\mathbf{y}_{corrected} = a$
        \Statex \textbf{else}:
        \Statex \quad $\mathbf{y}_{corrected} = \mathbf{y}_{pred}$
        \State \Return corrected label $\mathbf{y}_{corrected}$.
        
        \end{algorithmic}  
    \end{algorithm}

    The error corrector takes advantage of the "blessing of dimensionality" through linear separation, which demands minimal computations. This efficiency allows it to operate within a remarkably short time frame (less than 1 ms) and with low power requirements. Consequently, it can be effectively supported using a low-power printed circuit board (PCB) paired with a modest-sized battery (900 mAh). These attributes make the error corrector an excellent fit for low-powered devices.


\section{Hardware and Data}\label{sec:Hardware}
    \subsection{Capacitive sensor controller}
        
\begin{figure*}[t]
            \centering
            \centerline{\includegraphics[width=1\textwidth]{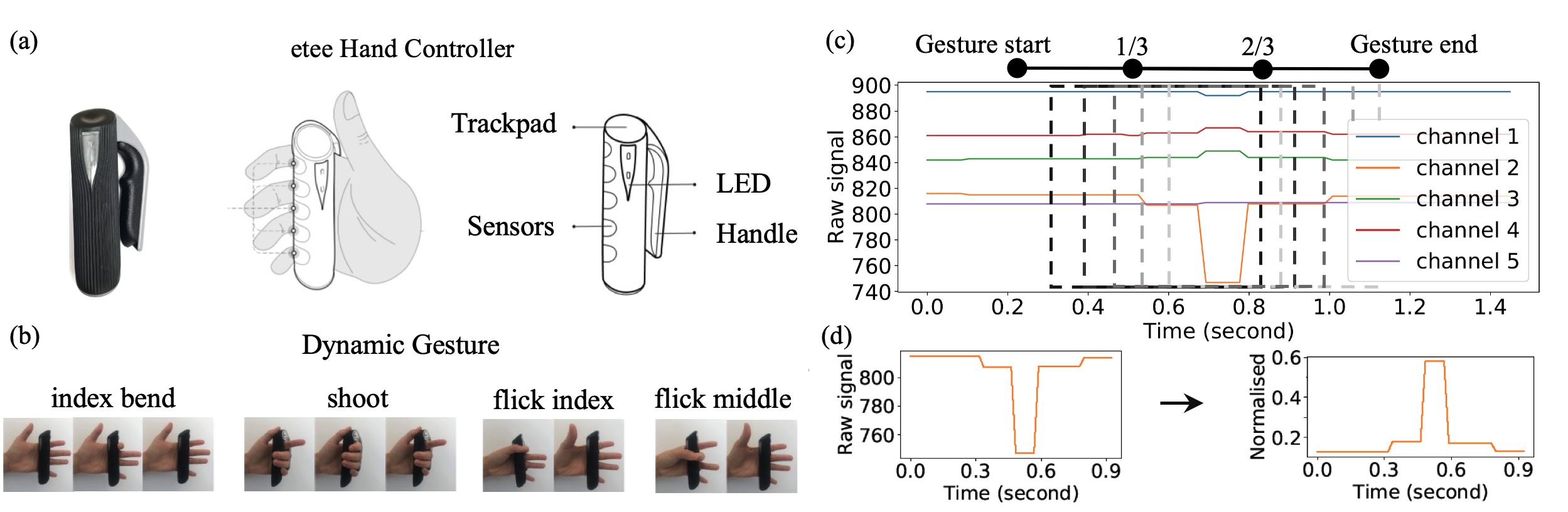}}
            \caption{\textbf{(a)} The wearable {\it etee} hand controllers used for collecting gesture recognition data in this work. The controller has a cylindrical silicon shell that fits in the palm of the users' hand. The transparent skeleton supports the rigid structure of the controller. The black sensors wrap around the skeleton and detect signals from each finger as they move and interact with the controller. Inside the skeleton, there is a PCB with a MCU and a battery to support all functions.  \textbf{(b)} The names and movements of the 4 dynamic gestures collected for this study. \textbf{(c)} During data collection, gesture start and end were marked. A 500 ms time window, represented by the dashed box, was applied to the signal to extract segments, sliding from the black box to the grey box with each signal frame. \textbf{(d,e)} Original signals on the left are normalised between 0 and 1. The normalisation range is defined by each user and sensor, with zero meaning that the fingers are fully open and one indicating that the sensor is being applied with full pressure by the hand.}
            \label{fig:hardware}
        \end{figure*}

        The controller hardware is composed of a capacitive touch sensor fusion unit, a printed circuit board (PCB) specifically designed for a microcontroller unit (ESP32), and supplementary components, including LEDs, a battery, and others, as illustrated in Figure~\ref{fig:hardware}(a). Detailed specifications of this hardware setup are available on the \textit{etee} website \cite{etee}. The sensors are strategically placed between the outer silicon shell and the skeleton of the controller. These sensors capture signals related to finger proximity, touch, and pressure across the area where a hand grips the controller. One of these sensors is positioned on the top of the controller, adjacent to the LED, while the other four rectangular sensors are wrapped around the cylindrical skeleton. Each sensor is aligned with a specific finger area and is sensitive to the capacitive signal changes caused by finger movements.
        
        The silicon shell surrounding the sensors serves multiple purposes. It acts as an insulating layer, enhancing the stability of the sensors' signals against direct skin contact. It also help modulate the strength of the signal in response to finger pressure. Moreover, the silicon shell's high elasticity provides a protective barrier for the rigid skeleton and the delicate electronics housed within, safeguarding them against external impacts and wear.
    
    \subsection{Dataset}\label{sec:Dataset}
        
        This section describes the methods for data collection and preprocessing for a gesture recognition classification task. We recruited twelve users to collect data on four hand gestures: "index bend", "shoot", "flick index", and "flick middle". Figure~\ref{fig:hardware}b illustrates the gestures and a series of corresponding movements. Each recording captured five time-series signals from five sensors, each correlating to a specific finger's movement. Figure~\ref{fig:hardware}c showcases the signal variation during the "index bend" gesture, where the index finger bends and then straightens, causing a distinct change in the signal at channel 2 (index finger sensor). The other sensors display minimal variation during this action. We extracted the dataset in a sliding window, where a 500 ms continuous signal segment is extracted from two-thirds of the gesture to its end, marked by the dashed box transitioning from dark to light in Figure~\ref{fig:hardware}c. 
        
        Additionally, we recorded signals for four dynamic gestures and extracted data for a "none" gesture, characterised by random hand movements that do not conform to a specific gesture pattern. This "none" gesture data was labelled from the end to the start of a gesture. Following data extraction, we normalised the signals as demonstrated in Figure~\ref{fig:hardware}d. The original signals were normalised to a range of [0,1], with zero representing a fully open finger and one indicating maximum pressure applied by the finger on the sensor.

        Each sample in the exact dataset, denoted as $\mathbf{x}$, has a data shape of (5, $t$), representing time-series signals from five sensors over $t$ time stamps. We noted that the average duration for a complete gesture is approximately 500 ms. Consequently, we set $t = 20$ when the hardware's transmitting frequency was 40 Hz. To maintain consistency with this gesture duration, the time window in the sliding dataset was also set to 500 ms. The dataset was then flattened into 100 features, with the first and second sets of 20 features representing the thumb and index finger signals, respectively.

        The signal patterns for each gesture vary across users due to differences in hand size and movement habits. Therefore, the dataset is categorised by the users, and further divided into training and testing sets. Each set contains signals from different users.


\section{Results and Discussion}\label{sec:results}
    
    This section presents the base model development of gesture recognition system, the prediction error analysis and the adaptive error corrector system performance evaluation.

    \subsection{Dimensionality Reduction}

\begin{figure}[t]
            \centerline{\includegraphics[width=0.8\textwidth]{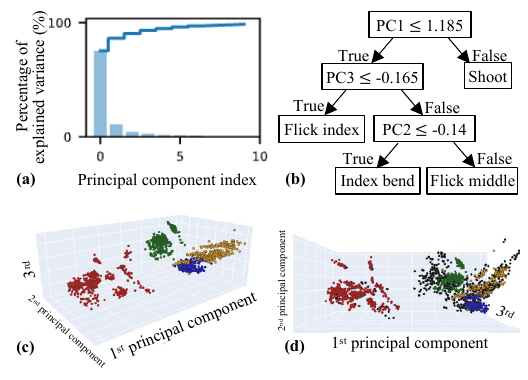}}
            \caption{\textbf{(a)} The solid line is the accumulated percentage of explained variance and the bar is the percentage. The first three PCs cover over 95\% of the explained variance in the dataset. \textbf{(b)} All 100 PCs were fed to a decision tree for classification of 4 main dynamic gestures. Only three, first PC (PC1), second PC (PC2) and third PC (PC3) were required for the decision making. \textbf{(c)} Three features - the top three PCs -  were used to visualise the dataset and show great sparsity among four gesture labels (colour represents the gesture here). \textbf{(d)} An extra "none" label (black) were added in the dataset showing that they are siting around all the other four gestures.}
            \label{fig:pca_analysis}
        \end{figure}

        We employed PCA to reduce the number of informative attributes entering the pattern recognition pipeline through scikit-learn's built-in function \cite{scikit}. The first three principal components (PCs) accounted for over 95\% of the variance from the original 100 features (Figure~\ref{fig:pca_analysis}). Additionally, a decision tree-based classification, using the 100 transformed PCs, identified the first three PCs as the most effective for splitting the data. These first three PCs  presented already reveals distinct separations in the data points among different gestures. However, we observed that some finger movements, not registered as gestures, appeared as data points close to those of actual gesture data (Figure~\ref{fig:pca_analysis}d). The inclusion of these "None" gestures—random hand movements not fitting a defined gesture pattern indicates that using only three PCs might be insufficient to fully differentiate between gesture and non-gesture data.

    \subsection{Base Model's Performance}

\begin{figure}[h]
                \centerline{\includegraphics[width=0.8\textwidth]{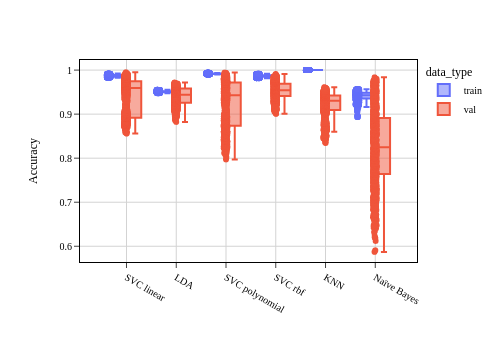}}
                \caption{The accuracy box plot of $k$-fold cross-validation for six base systems. The dataset were randomly shuffled and split to the train set and validation set in groups of users. All base systems showed accuracy with train set are close to 1 while the validation accuracy varies around 0.9.}
                \label{fig:base_model_accuracy}
            \end{figure}

        We evaluated six base systems: K-Nearest Neighbours (KNN), SVM with linear, polynomial, and radial basis function (RBF) kernels, LDA, and Naïve Bayes (NB). Each system consisted of a combination of dimension reduction using PCA and classification. Figure~\ref{fig:base_model_accuracy} displays the $k$-fold cross-validation results, highlighting the best performance for each system across various hyper-parameter combinations. Each marker in Figure~\ref{fig:base_model_accuracy} represents an accuracy result from a training or validation dataset. These datasets were randomly split in group of users with an 8:3 ratio within the full dataset. In this setup, the classifier selected several PCs as inputs to differentiate gestures. The Linear SVM classifier and the NB model utilised all 100 PCs. The Polynomial SVM classifier achieved the highest average validation accuracy using 20 PCs. The RBF-SVM and LDA yielded the best results with the top 10 PCs. The KNN model reached an average validation accuracy exceeding 0.9 with 5 PCs. The box plot depicting the accuracy of both the training and validation sets indicates that the RBF-SVM model with 10 PCs outperforms all other models in validation accuracy. This model achieves an average accuracy above 0.95, maintaining minimal variation compared to the other models.
       
        Subsequently, we introduced a dataset from a new user, who had not been involved in the analysis, train and test stage of the base model. The performance of the trained base model on this new dataset showed a significant decrease, primarily due to the distinctly different hand movement patterns of the new user (see Table.~\ref{tab:base_model_accuracy}). However, when we retrained the base model using the training dataset from this new user, it demonstrated excellent gesture prediction accuracy for that specific user. Unfortunately, this resulted in a loss of generalised predictive performance for other users.
        
        \begin{table}[h]
            \caption{Base model accuracy (\%)}
            \begin{center}
            \resizebox{1\columnwidth}{!}{\begin{tabular}{lcccc}
                \multicolumn{1}{c}{\multirow{2}{*}{SVM}} & \multicolumn{4}{c}{Test on}                                                     \\ \cline{2-5} 
                \multicolumn{1}{c}{}                     & Base Model Trainset & Base Model Testset & New User Trainset & New User Testset \\ \hline
                Trained on Base Model Trainset           & 98.3                & 96.3               & 87.7              & 80.4             \\
                Trained on New User Trainset             & 84                  & 92                 & 99.6              & 99.2             \\ \hline
                \end{tabular}}
            \label{tab:base_model_accuracy}
            \end{center}
        \end{table}
 
        To address the challenges encountered, we incorporated the error correction system into the existing gesture recognition framework (Figure~\ref{fig:workflow}). This system is a hybrid of the base model, which handles initial gesture recognition, and an additional component known as the corrector. The corrector's role is to refine and adjust the outputs of the base model, thereby adapting the model performance for customised user, especially in cases where the base model may struggle to generalise for a new user. This integrated approach aims to maintain high performance levels across diverse user dataset, while also customising the gesture recognition model to cater to individual user needs.

    \subsection{Error analysis} 
\begin{figure*}
            \centerline{\includegraphics[width=1\textwidth]{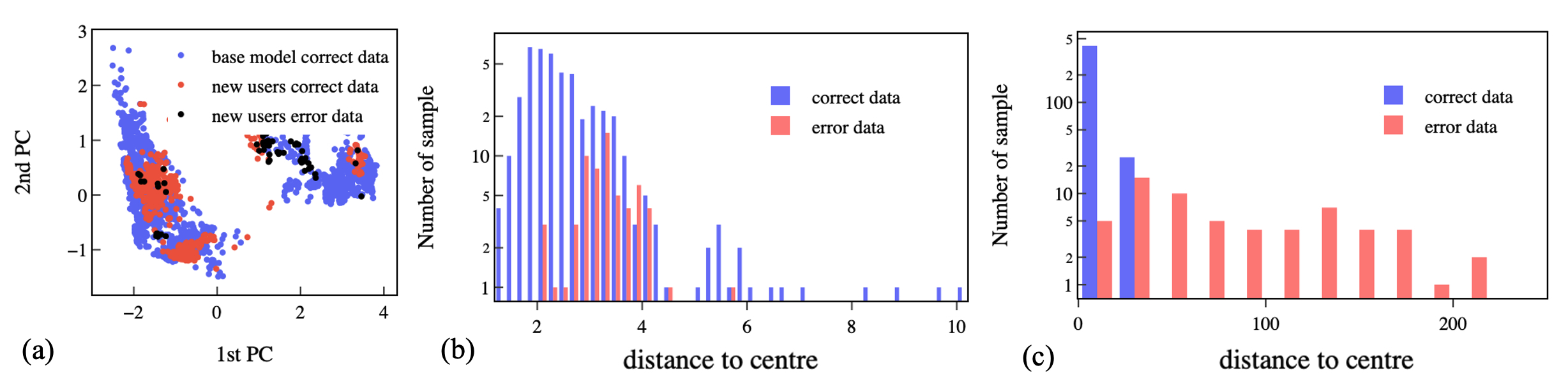}}
            \caption{The correct samples and error samples are separable thorough Euclidean distance value to the correct data centre. (a) Correct samples and error samples are overlaying with each other from the first two PCs with the highest eigen value. (b) 8 PCs were selected here to calculated the Euclidean distance value of the dataset to the centre of the correct data with 8 features. The distance of the error sample are in the same range as the correct samples. (c) The 8 PCs were expanded through polynomial kernel map with degree of 5. The Euclidean distance calculated in the high dimensional feature space shows that correct data (less than 22) are separable from the errors (mostly above 40). }
            \label{fig:error_analysis}
        \end{figure*}

        The dataset from the new user was divided into two categories: a 'correct' set, where the base model's predictions matched the ground truth of the gestures, and an 'error' set, where the predictions were incorrect. Figure~\ref{fig:error_analysis}(a) shows the scattered plot of  data points from the first two PCs and we observe that the 'correct' portion of the new dataset overlapped with the base model's training dataset. The 'error' set, obtained from the same new user, exhibited a scatter pattern similar to the 'correct' set, making them indistinguishable based solely on the first two PCs. Subsequently, we selected the first 8 PCs as an example feature space and identified the centre of the 'correct' set. We then calculated the Euclidean distance from both sets to this centre. The histogram in Figure~\ref{fig:error_analysis}(b) illustrates that errors are not readily distinguishable from the Euclidean distance values alone, as there is a significant overlap in the distribution of distances to the centre between the 'correct' and 'error' sets.

        To differentiate the 'error' set from the 'correct' set using a linear separator, we projected the features into a higher-dimensional feature space, as described in \cite{corrector_application2021}. This was achieved using a polynomial kernel of degree 5 to map the 8 PCs onto 859 features. The features underwent whitening, following the procedure outlined in Algorithm~\ref{algo: train}. Figure~\ref{fig:error_analysis}(c) displays the histogram of distances from the 'correct' set to the centre. Most samples from the 'correct' set are concentrated within a range of [10, 22], while most of the errors are situated further from the centre, exceeding the distance of 40. Therefore, they are separable from the 'correct' set using a linear separator with an acceptable error rate.
    
    \subsection{Error Corrector}
    \label{subsec:ec}
       
        Following the error analysis of the new user data in conjunction with the base model, we investigated the use of polynomial kernels with degrees ranging from 2 to 9. This was done in combination with the first 8 PCs to project the data into a high-dimensional feature space before applying the error corrector classifier. The dataset from the new user, represented in this high-dimensional feature space, was then subjected to classification using a Centroid classifier. This classifier was tasked with distinguishing the 'error' set from the 'correct' set. To achieve this, we employed a threshold parameter within the classifier, which determined the boundary for this binary classification. 
        
        To this end, we introduced another classifier to categorise the error samples by error type using LDA, to identify the specific type of prediction error to which each sample was most susceptible. This distinction was crucial, as a gesture predicted as one type when it should have been another represents a different error type. For example, gesture A predicted as gesture B constitutes a different error type than gesture A predicted as gesture C. If the predicted label of an erroneous sample matched the pattern of a specific error type, we could then rectify the outcome of the gesture recognition base model. 

        Ten distinct feature types were analysed after training within the error correction system. This analysis encompassed the first 8 PCs from the base model's PCA transformation, which were utilised as a benchmark, and features mapped through polynomial kernel transformations with degrees ranging from 1 to 9. These features were then processed through a whitening procedure. The centroid classifier employed a threshold $\Delta$ to effectively differentiate between accurate predictions and errors.

        \begin{figure*}[t]
            \centerline{\includegraphics[width=1\textwidth]{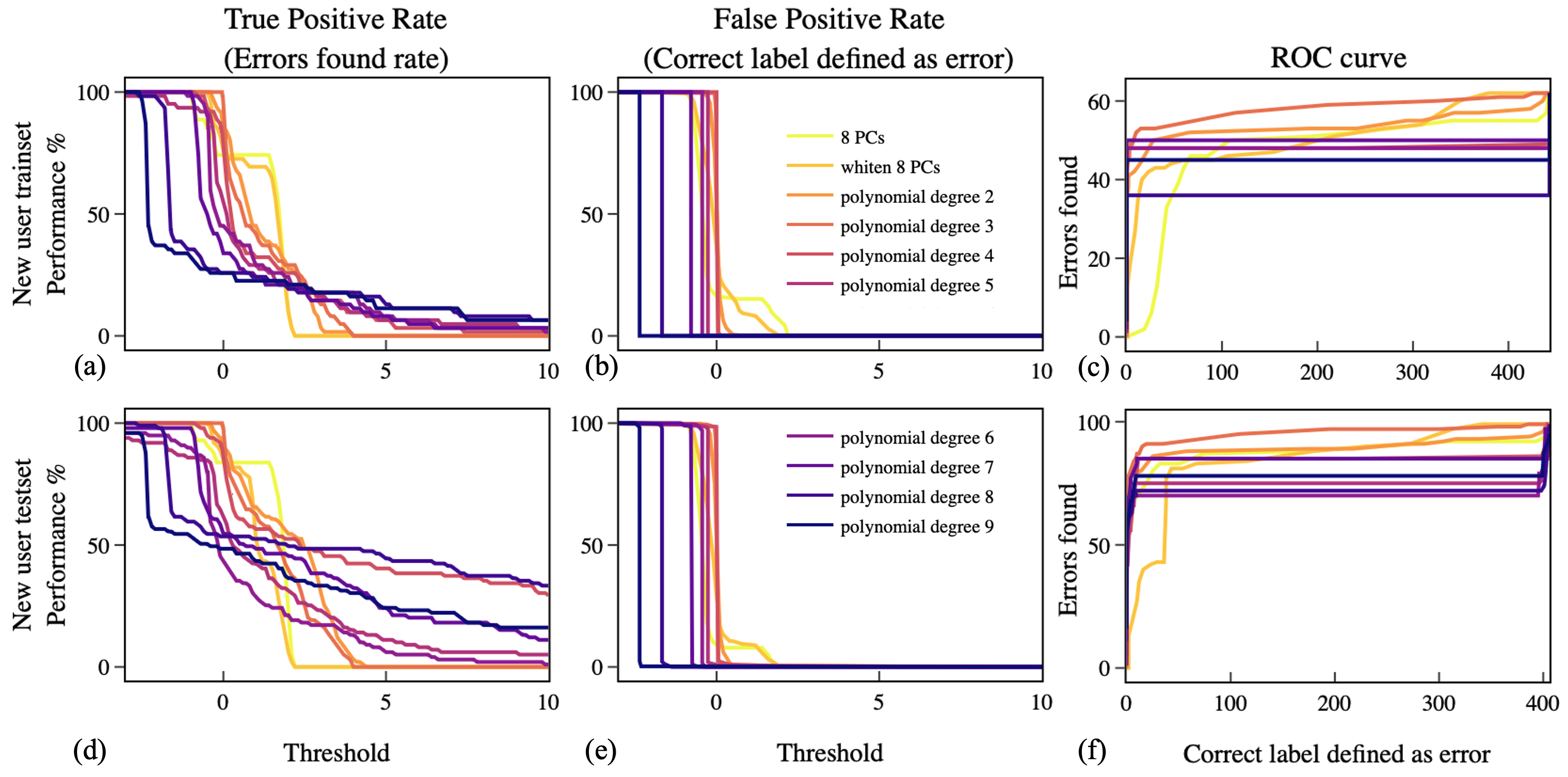}}
            \caption{(a) The percentage of errors found out of all errors (TPR) as a function of threshold in the error corrector in the training set from a new user. (b) The percentage of correct label defined as error (FPR) as a function of threshold in the training set from a new user. (c) The ROC curve with the TP number (errors found), and FP number (correct label defined as error) for the training set. (d) The TPR as a function of threshold in the testing set. (e) The FPR as a function of threshold in the testing set from a new user. (f) The ROC curve for the testing set.}
            \label{fig:roc}
        \end{figure*}

        Figure~\ref{fig:roc} demonstrates the True Positive Rate (TPR), False Positive Rate (FPR) and Receiver Operating Characteristic (ROC) curve as functions of the threshold used in the error corrector. From the TPR plot in  Figure~\ref{fig:roc}a and  Figure~\ref{fig:roc}d, we observed that the TRP declines slower as the dimension of the feature space increases. The FPR, on the other hand, has a sharper drop as we rose the threshold under high dimension cases (Figure~\ref{fig:roc}b and Figure~\ref{fig:roc}d). It suggests that the gap between the correct set and the error set is bigger when the dimension of feature space is high. This is also confirmed from the ROC curve in  Figure~\ref{fig:roc}c and Figure~\ref{fig:roc}f. Under high dimensional feature space when polynomial kernel degree is above 5, we can find a threshold with optimised performance, where errors are identified without touching the correct labels.
        
\begin{figure*}[t]
            \centerline{\includegraphics[width=1\textwidth]{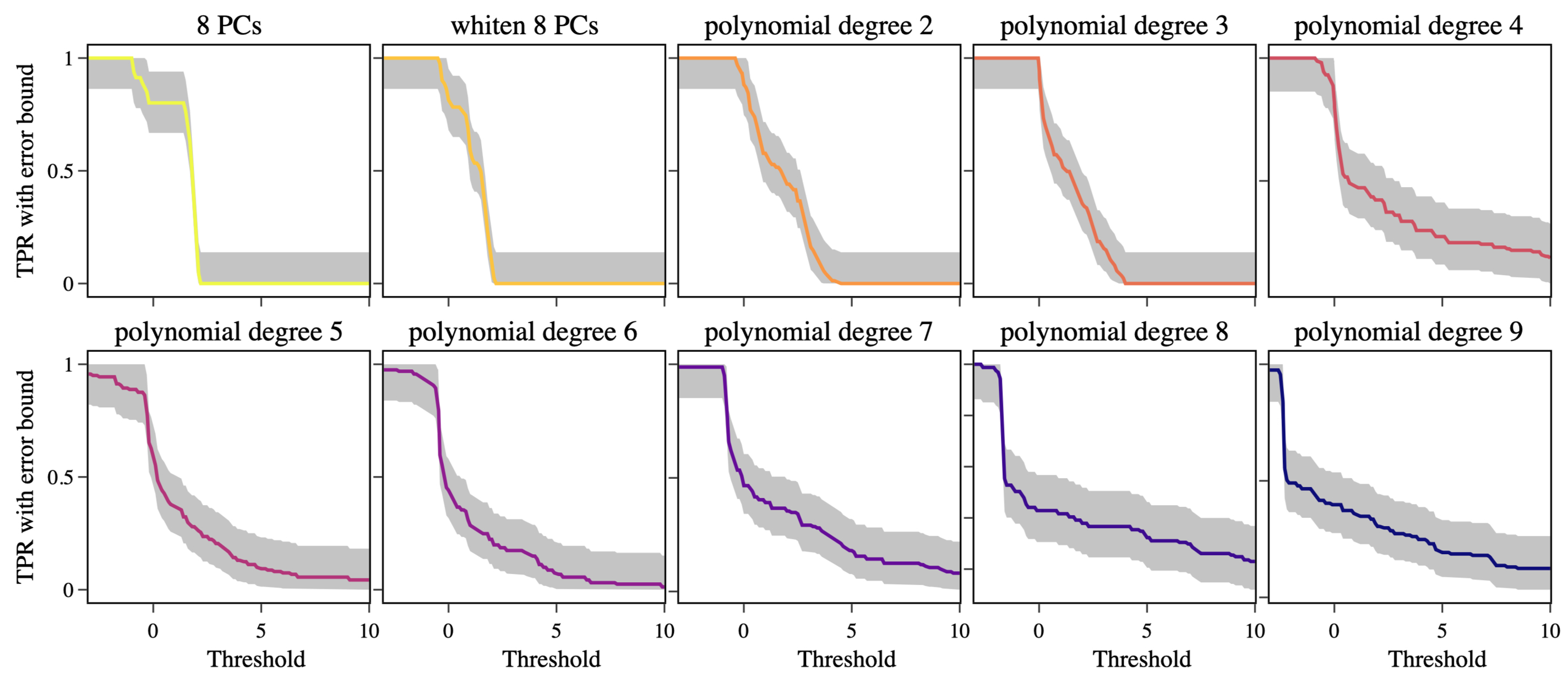}}
            \caption{The lower bound and upper bound on the probability of the correct rejection of errors as a function of the threshold.}
            \label{fig:error_bound}
        \end{figure*}

        According to the recent work on the quantification of errors in AI corrector \cite{Tyukin2024}, the probability of the correct rejection of errors is bounded with the threshold value and the error dataset size. Figure~\ref{fig:error_bound} presents the error bounds as a function of the threshold, with the error dataset size of 161 within both the new user training set and testing set. The correct rejection rate of errors does not exceed the shaded area based on the new user's data.

\begin{figure*}
            \centerline{\includegraphics[width=1\textwidth]{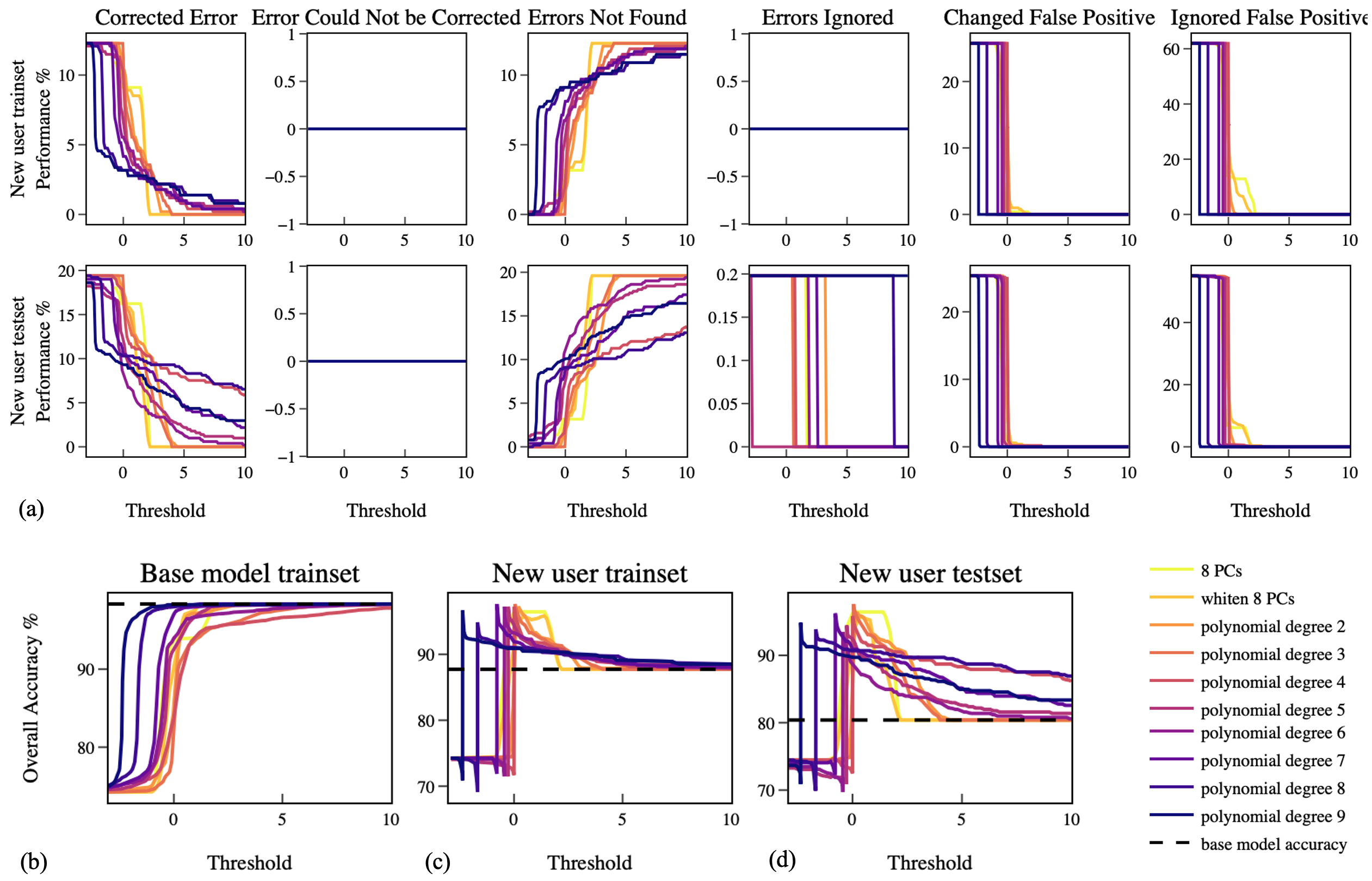}}
            \caption{(a) The first row displays the performance of the error corrector system on the new user training set. The second row illustrates the performance on the new user test set. (b) The performance of the error correction system on the original base model training set suggests that error corrector would reduce the performance under low dimension kernel map but remain it's performance with high dimension kernel map. (c) The overall accuracy of error correction system with the new user training dataset has an overall improvement compared with the base model itself. (d) The accuracy of error correction system applied on the new user testing dataset confirms the improvement of prediction accuracy regardless of the kernel map degree.}
            \label{fig:corrector_performance}
        \end{figure*}

        After errors are identified, we need to correct them to the right labels. Figure~\ref{fig:corrector_performance} presents a performance chart for the centroid classifier and error-type classifier, examining its output relative to various threshold values. The error correction system was assessed by the following events:
         \begin{itemize}
             \item Corrected Error: This event occurs when an error is accurately detected and it aligns with the identified error type pattern, leading to its successful correction.
             \item Error Could Not be Corrected: Although an error is detected and matches an error type pattern, the true label does not correspond to the indicated error type, resulting in a persistently incorrect gesture classification.
             \item Errors Not Found: Errors are misclassified as correct predictions.
             \item Error Ignored: Errors are identified but do not conform to any error type pattern and are therefore disregarded.
             \item Changed FP: A correct prediction is incorrectly labelled as an error and, as a result, is assigned a different output label.
             \item Ignored False Positive: Correct predictions mislabelled as errors; however, these do not match any error type pattern and are thus ignored.
             \item Improved Overall Accuracy: The comprehensive enhancement in accuracy achieved by the error correction system on the new user dataset, in comparison to the performance of the base model.
         \end{itemize}

        Figure~\ref{fig:corrector_performance}a demonstrates that the new training and testing sets displayed a consistent performance pattern. For features that did not undergo polynomial kernel expansion, a notable difference was observed compared to the remaining of the feature maps. Regarding features expanded with polynomial kernels, we noticed a gradual decline in the corrected error plots for both the new training and testing sets as the threshold increased, indicating a stricter criterion for error determination. This is consistent with the observation in the ROC curve.  

        From the data in Figure~\ref{fig:corrector_performance}a, it's evident that the overall improved performance of the error correction system was largely influenced by only two factors in this case: the corrected errors and the changed FP. A positive corrected error performance would improve overall accuracy compared with the base model, while a positive changed FP performance would decrease it. 

        When we applied the corrector to the base model's training set (Figure~\ref{fig:corrector_performance}b), we observed that all feature spaces had an overall negative impact on the base model's performance. However, this negative impact can be restricted in a smaller range of threshold when polynomial kernel degree is high. In contrast, feature maps with a small polynomial degree remain a negative impact on the base model performance. Figure~\ref{fig:corrector_performance}c and \ref{fig:corrector_performance}d show the gesture recognition accuracy after error correction system with both train set and test set from a new user, compared with the base model, the error correction system exhibit obvious improvement in the pattern recognition performance, regardless of the polynomial degree in the high-dimensional kernel map. When the dimension of the feature space is higher, there is a better distinguish between errors and correct set, offering a positive overall accuracy for new user's data in a larger threshold range. Beyond these threshold selection, the performance of this error correction system decreases. These observations suggest that while the error corrector is beneficial in enhancing gesture recognition accuracy, there is a balance to be struck between customisation and maintaining the performance of the base model.

         In combination with the error correction performance from the base model train set, we selected a polynomial kernel of degree 9 with a threshold of 0.2 to achieve maximum accuracy in error detection. The overall performance of the error correction system is then improving for a customised user, as well as maintaining the base model performance.  According to the quantification of errors in AI corrector, the probability of the correct rejection of errors with the value threshold equalling 0.2 and with the size of the error dataset used to produce the corrector of 161 data points is higher than 27.6\% and lower than 53.6\%. The inconsistency is likely due to the small volume of our new user's dataset.
        
        The observed changes in accuracy on the new user data  led us to explore the intrinsic dimension of the training dataset. This concept is a significant one in modern machine learning and has garnered widespread discussion. The intrinsic dimension of a dataset, as pointed out, is not always indicative of the number of features it possesses \cite{stochastic2021}. For instance, a dataset comprising three features but distributed on a 2D plane effectively has an intrinsic dimension of 2, not 3. There are various definitions of intrinsic dimension, and for our analysis, we adopted the Fisher separability statistic-based dimensionality proposed by Gorban and Tyukin \cite{error_corrector_2018}. This definition aligns with PCA-based measurements of intrinsic dimension and is adept at capturing low-dimensional fractal and fine-grained structures within data.
        
        \begin{figure*}
            \centerline{\includegraphics[width=1\textwidth]{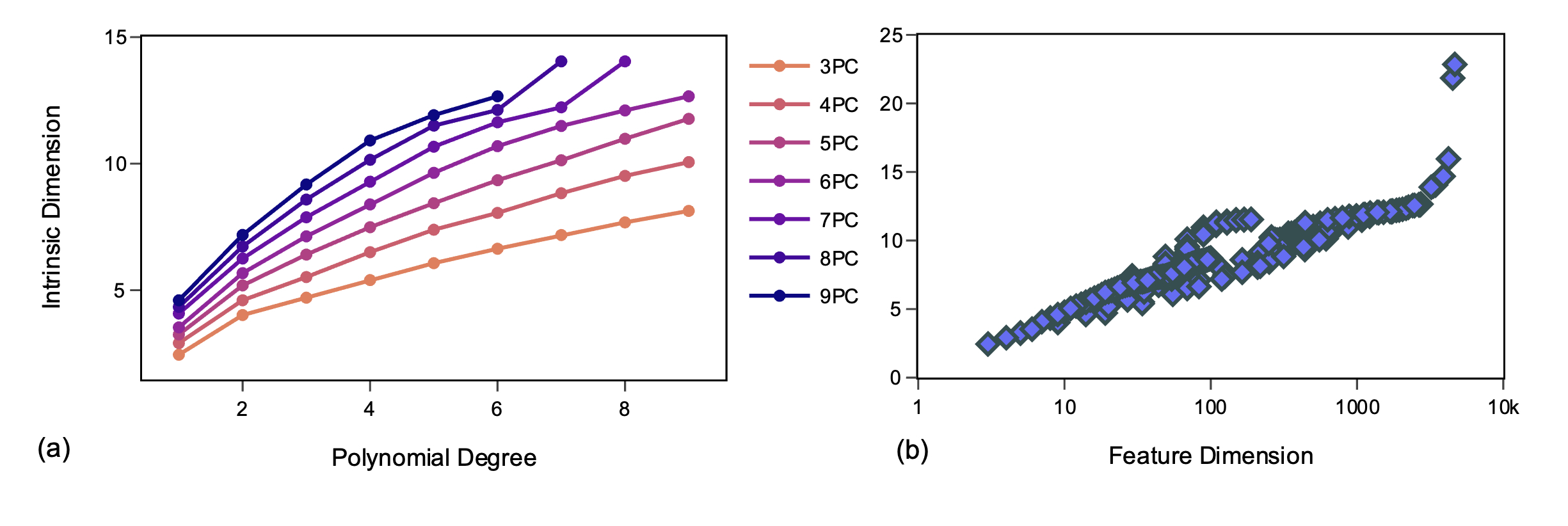}}
            \caption{(a)The intrinsic dimension of features projected using the polynomial kernel map. The original features ranged from 3 to 9 PCs. Each feature set underwent transformations using polynomial kernels with degrees varying from 1 to 9. Notably, when the number of features exceeds the number of samples in the dataset, the intrinsic dimension tends to approach infinity. (b) This part explores the relationship between the intrinsic dimension and the number of features after applying kernel tricks. There is a noticeable increase in intrinsic dimension as the number of features rises. Crucially, when the number of features approaches the number of samples, the calculated intrinsic dimension spikes sharply towards infinity. In this context, the sample size is approximately 5000.}
            \label{fig:intrinsic_dimension}
        \end{figure*}
        
        When the feature count in our dataset surpasses a certain threshold, the intrinsic dimension becomes challenging to calculate accurately, mainly due to the limited sample size. Nonetheless, as illustrated in Figure~\ref{fig:intrinsic_dimension}, the trend of the intrinsic dimension remains observable with the escalation of specific parameter values. The highest measurable intrinsic dimension noted did not exceed 12, aligning approximately with the intrinsic dimension of the original input data. To augment the dataset's dimensionality, we merged features derived from two different kernel transformations to create a new feature space. This combined set of features was then applied to subsequent classifiers for the categorisation of error groups and the implementation of error correction.

\section{Conclusions}\label{sec:conclusion}

    In this study, we have introduced a hand gesture recognition system that integrates a straightforward base model with an error corrector, designed to adapt the gesture recognition model for customised users while preserving generalised performance. We tested this system on low-dimensional capacitive sensor signals measured by the \textit{etee} hand controller. We compared the performance of the PCA and a combination of six classifiers in the base model, selecting the top performer. Although the best classifier occasionally exhibited errors, its performance declined when applied to datasets collected from new users. To counter these occasional errors, we incorporated an adaptive error correction mechanism into the system, where the base model is accompanied by a corrector. The results showed an enhancement in overall accuracy and the maintenance of robust performance for the training dataset, provided the threshold and kernel map are judiciously chosen.



\bibliography{references}

\end{document}